\documentclass[12pt]{article}
\usepackage{amsmath,amssymb}
\usepackage{graphicx,wrapfig}
\usepackage{cite,color}
 \allowdisplaybreaks
\newcommand{\be}{\begin{equation}}
\newcommand{\ee}{\end{equation}}
\newcommand{\bea}{\begin{eqnarray}}
\newcommand{\eea}{\end{eqnarray}}

\newcommand{\bx}{{\bf x}}

\newcommand{\cP}{{\cal P}}
\newcommand{\cQ}{{\cal Q}}
\newcommand{\cR}{{\cal R}}

\newcommand{\cZ}{{\cal Z}}

\newcommand{\ds}{\displaystyle}
\newcommand{\nn}{\nonumber}
\newcommand{\pd}{\partial}
\newcommand{\td}{\tilde}

\newcommand{\one}{{\bf 1}}

\def\brk#1{\left\langle#1\right\rangle}

\long\def\symbolfootnote[#1]#2{\begingroup%
\def\thefootnote{\fnsymbol{footnote}}\footnote[#1]{#2}\endgroup}

\newcommand{\ga}{\alpha}
\newcommand{\gb}{\beta}

\newcommand{\TwistorialWitten}{
\ifx\JPicScale\undefined\def\JPicScale{0.8}\fi \unitlength
\JPicScale mm
\begin{picture}(40.26,40.76)(-5,30)
\linethickness{0.2mm}
\put(18,53){$x$}\put(8,84){$x_1$}\put(40,70){$x_2$}\put(-10,60){$x_n$}
\put(40.25,61){\line(0,1){0.5}} \put(19,61){\circle*{5}}
\multiput(40.24,62)(0.01,-0.5){1}{\line(0,-1){0.5}}
\multiput(40.22,62.49)(0.02,-0.5){1}{\line(0,-1){0.5}}
\multiput(40.19,62.99)(0.03,-0.5){1}{\line(0,-1){0.5}}
\multiput(40.14,63.48)(0.05,-0.49){1}{\line(0,-1){0.49}}
\multiput(40.08,63.97)(0.06,-0.49){1}{\line(0,-1){0.49}}
\multiput(40.01,64.47)(0.07,-0.49){1}{\line(0,-1){0.49}}
\multiput(39.93,64.96)(0.08,-0.49){1}{\line(0,-1){0.49}}
\multiput(39.84,65.45)(0.09,-0.49){1}{\line(0,-1){0.49}}
\multiput(39.74,65.93)(0.1,-0.49){1}{\line(0,-1){0.49}}
\multiput(39.63,66.42)(0.11,-0.48){1}{\line(0,-1){0.48}}
\multiput(39.5,66.9)(0.12,-0.48){1}{\line(0,-1){0.48}}
\multiput(39.37,67.37)(0.14,-0.48){1}{\line(0,-1){0.48}}
\multiput(39.22,67.85)(0.15,-0.47){1}{\line(0,-1){0.47}}
\multiput(39.06,68.32)(0.16,-0.47){1}{\line(0,-1){0.47}}
\multiput(38.89,68.79)(0.17,-0.47){1}{\line(0,-1){0.47}}
\multiput(38.71,69.25)(0.18,-0.46){1}{\line(0,-1){0.46}}
\multiput(38.52,69.71)(0.1,-0.23){2}{\line(0,-1){0.23}}
\multiput(38.32,70.16)(0.1,-0.23){2}{\line(0,-1){0.23}}
\multiput(38.11,70.61)(0.11,-0.22){2}{\line(0,-1){0.22}}
\multiput(37.89,71.06)(0.11,-0.22){2}{\line(0,-1){0.22}}
\multiput(37.66,71.5)(0.12,-0.22){2}{\line(0,-1){0.22}}
\multiput(37.42,71.93)(0.12,-0.22){2}{\line(0,-1){0.22}}
\multiput(37.16,72.36)(0.13,-0.21){2}{\line(0,-1){0.21}}
\multiput(36.9,72.78)(0.13,-0.21){2}{\line(0,-1){0.21}}
\multiput(36.63,73.2)(0.14,-0.21){2}{\line(0,-1){0.21}}
\multiput(36.35,73.61)(0.14,-0.2){2}{\line(0,-1){0.2}}
\multiput(36.06,74.01)(0.15,-0.2){2}{\line(0,-1){0.2}}
\multiput(35.76,74.41)(0.15,-0.2){2}{\line(0,-1){0.2}}
\multiput(35.45,74.8)(0.1,-0.13){3}{\line(0,-1){0.13}}
\multiput(35.14,75.18)(0.11,-0.13){3}{\line(0,-1){0.13}}
\multiput(34.81,75.55)(0.11,-0.12){3}{\line(0,-1){0.12}}
\multiput(34.47,75.92)(0.11,-0.12){3}{\line(0,-1){0.12}}
\multiput(34.13,76.28)(0.11,-0.12){3}{\line(0,-1){0.12}}
\multiput(33.78,76.63)(0.12,-0.12){3}{\line(1,0){0.12}}
\multiput(33.42,76.97)(0.12,-0.11){3}{\line(1,0){0.12}}
\multiput(33.05,77.31)(0.12,-0.11){3}{\line(1,0){0.12}}
\multiput(32.68,77.64)(0.12,-0.11){3}{\line(1,0){0.12}}
\multiput(32.3,77.95)(0.13,-0.11){3}{\line(1,0){0.13}}
\multiput(31.91,78.26)(0.13,-0.1){3}{\line(1,0){0.13}}
\multiput(31.51,78.56)(0.2,-0.15){2}{\line(1,0){0.2}}
\multiput(31.11,78.85)(0.2,-0.15){2}{\line(1,0){0.2}}
\multiput(30.7,79.13)(0.2,-0.14){2}{\line(1,0){0.2}}
\multiput(30.28,79.4)(0.21,-0.14){2}{\line(1,0){0.21}}
\multiput(29.86,79.66)(0.21,-0.13){2}{\line(1,0){0.21}}
\multiput(29.43,79.92)(0.21,-0.13){2}{\line(1,0){0.21}}
\multiput(29,80.16)(0.22,-0.12){2}{\line(1,0){0.22}}
\multiput(28.56,80.39)(0.22,-0.12){2}{\line(1,0){0.22}}
\multiput(28.11,80.61)(0.22,-0.11){2}{\line(1,0){0.22}}
\multiput(27.66,80.82)(0.22,-0.11){2}{\line(1,0){0.22}}
\multiput(27.21,81.02)(0.23,-0.1){2}{\line(1,0){0.23}}
\multiput(26.75,81.21)(0.23,-0.1){2}{\line(1,0){0.23}}
\multiput(26.29,81.39)(0.46,-0.18){1}{\line(1,0){0.46}}
\multiput(25.82,81.56)(0.47,-0.17){1}{\line(1,0){0.47}}
\multiput(25.35,81.72)(0.47,-0.16){1}{\line(1,0){0.47}}
\multiput(24.87,81.87)(0.47,-0.15){1}{\line(1,0){0.47}}
\multiput(24.4,82)(0.48,-0.14){1}{\line(1,0){0.48}}
\multiput(23.92,82.13)(0.48,-0.12){1}{\line(1,0){0.48}}
\multiput(23.43,82.24)(0.48,-0.11){1}{\line(1,0){0.48}}
\multiput(22.95,82.34)(0.49,-0.1){1}{\line(1,0){0.49}}
\multiput(22.46,82.43)(0.49,-0.09){1}{\line(1,0){0.49}}
\multiput(21.97,82.51)(0.49,-0.08){1}{\line(1,0){0.49}}
\multiput(21.47,82.58)(0.49,-0.07){1}{\line(1,0){0.49}}
\multiput(20.98,82.64)(0.49,-0.06){1}{\line(1,0){0.49}}
\multiput(20.49,82.69)(0.49,-0.05){1}{\line(1,0){0.49}}
\multiput(19.99,82.72)(0.5,-0.03){1}{\line(1,0){0.5}}
\multiput(19.5,82.74)(0.5,-0.02){1}{\line(1,0){0.5}}
\multiput(19,82.75)(0.5,-0.01){1}{\line(1,0){0.5}}
\put(18.5,82.75){\line(1,0){0.5}}
\multiput(18,82.74)(0.5,0.01){1}{\line(1,0){0.5}}
\multiput(17.51,82.72)(0.5,0.02){1}{\line(1,0){0.5}}
\multiput(17.01,82.69)(0.5,0.03){1}{\line(1,0){0.5}}
\multiput(16.52,82.64)(0.49,0.05){1}{\line(1,0){0.49}}
\multiput(16.03,82.58)(0.49,0.06){1}{\line(1,0){0.49}}
\multiput(15.53,82.51)(0.49,0.07){1}{\line(1,0){0.49}}
\multiput(15.04,82.43)(0.49,0.08){1}{\line(1,0){0.49}}
\multiput(14.55,82.34)(0.49,0.09){1}{\line(1,0){0.49}}
\multiput(14.07,82.24)(0.49,0.1){1}{\line(1,0){0.49}}
\multiput(13.58,82.13)(0.48,0.11){1}{\line(1,0){0.48}}
\multiput(13.1,82)(0.48,0.12){1}{\line(1,0){0.48}}
\multiput(12.63,81.87)(0.48,0.14){1}{\line(1,0){0.48}}
\multiput(12.15,81.72)(0.47,0.15){1}{\line(1,0){0.47}}
\multiput(11.68,81.56)(0.47,0.16){1}{\line(1,0){0.47}}
\multiput(11.21,81.39)(0.47,0.17){1}{\line(1,0){0.47}}
\multiput(10.75,81.21)(0.46,0.18){1}{\line(1,0){0.46}}
\multiput(10.29,81.02)(0.23,0.1){2}{\line(1,0){0.23}}
\multiput(9.84,80.82)(0.23,0.1){2}{\line(1,0){0.23}}
\multiput(9.39,80.61)(0.22,0.11){2}{\line(1,0){0.22}}
\multiput(8.94,80.39)(0.22,0.11){2}{\line(1,0){0.22}}
\multiput(8.5,80.16)(0.22,0.12){2}{\line(1,0){0.22}}
\multiput(8.07,79.92)(0.22,0.12){2}{\line(1,0){0.22}}
\multiput(7.64,79.66)(0.21,0.13){2}{\line(1,0){0.21}}
\multiput(7.22,79.4)(0.21,0.13){2}{\line(1,0){0.21}}
\multiput(6.8,79.13)(0.21,0.14){2}{\line(1,0){0.21}}
\multiput(6.39,78.85)(0.2,0.14){2}{\line(1,0){0.2}}
\multiput(5.99,78.56)(0.2,0.15){2}{\line(1,0){0.2}}
\multiput(5.59,78.26)(0.2,0.15){2}{\line(1,0){0.2}}
\multiput(5.2,77.95)(0.13,0.1){3}{\line(1,0){0.13}}
\multiput(4.82,77.64)(0.13,0.11){3}{\line(1,0){0.13}}
\multiput(4.45,77.31)(0.12,0.11){3}{\line(1,0){0.12}}
\multiput(4.08,76.97)(0.12,0.11){3}{\line(1,0){0.12}}
\multiput(3.72,76.63)(0.12,0.11){3}{\line(1,0){0.12}}
\multiput(3.37,76.28)(0.12,0.12){3}{\line(1,0){0.12}}
\multiput(3.03,75.92)(0.11,0.12){3}{\line(0,1){0.12}}
\multiput(2.69,75.55)(0.11,0.12){3}{\line(0,1){0.12}}
\multiput(2.36,75.18)(0.11,0.12){3}{\line(0,1){0.12}}
\multiput(2.05,74.8)(0.11,0.13){3}{\line(0,1){0.13}}
\multiput(1.74,74.41)(0.1,0.13){3}{\line(0,1){0.13}}
\multiput(1.44,74.01)(0.15,0.2){2}{\line(0,1){0.2}}
\multiput(1.15,73.61)(0.15,0.2){2}{\line(0,1){0.2}}
\multiput(0.87,73.2)(0.14,0.2){2}{\line(0,1){0.2}}
\multiput(0.6,72.78)(0.14,0.21){2}{\line(0,1){0.21}}
\multiput(0.34,72.36)(0.13,0.21){2}{\line(0,1){0.21}}
\multiput(0.08,71.93)(0.13,0.21){2}{\line(0,1){0.21}}
\multiput(-0.16,71.5)(0.12,0.22){2}{\line(0,1){0.22}}
\multiput(-0.39,71.06)(0.12,0.22){2}{\line(0,1){0.22}}
\multiput(-0.61,70.61)(0.11,0.22){2}{\line(0,1){0.22}}
\multiput(-0.82,70.16)(0.11,0.22){2}{\line(0,1){0.22}}
\multiput(-1.02,69.71)(0.1,0.23){2}{\line(0,1){0.23}}
\multiput(-1.21,69.25)(0.1,0.23){2}{\line(0,1){0.23}}
\multiput(-1.39,68.79)(0.18,0.46){1}{\line(0,1){0.46}}
\multiput(-1.56,68.32)(0.17,0.47){1}{\line(0,1){0.47}}
\multiput(-1.72,67.85)(0.16,0.47){1}{\line(0,1){0.47}}
\multiput(-1.87,67.37)(0.15,0.47){1}{\line(0,1){0.47}}
\multiput(-2,66.9)(0.14,0.48){1}{\line(0,1){0.48}}
\multiput(-2.13,66.42)(0.12,0.48){1}{\line(0,1){0.48}}
\multiput(-2.24,65.93)(0.11,0.48){1}{\line(0,1){0.48}}
\multiput(-2.34,65.45)(0.1,0.49){1}{\line(0,1){0.49}}
\multiput(-2.43,64.96)(0.09,0.49){1}{\line(0,1){0.49}}
\multiput(-2.51,64.47)(0.08,0.49){1}{\line(0,1){0.49}}
\multiput(-2.58,63.97)(0.07,0.49){1}{\line(0,1){0.49}}
\multiput(-2.64,63.48)(0.06,0.49){1}{\line(0,1){0.49}}
\multiput(-2.69,62.99)(0.05,0.49){1}{\line(0,1){0.49}}
\multiput(-2.72,62.49)(0.03,0.5){1}{\line(0,1){0.5}}
\multiput(-2.74,62)(0.02,0.5){1}{\line(0,1){0.5}}
\multiput(-2.75,61.5)(0.01,0.5){1}{\line(0,1){0.5}}
\put(-2.75,61){\line(0,1){0.5}}
\multiput(-2.75,61)(0.01,-0.5){1}{\line(0,-1){0.5}}
\multiput(-2.74,60.5)(0.02,-0.5){1}{\line(0,-1){0.5}}
\multiput(-2.72,60.01)(0.03,-0.5){1}{\line(0,-1){0.5}}
\multiput(-2.69,59.51)(0.05,-0.49){1}{\line(0,-1){0.49}}
\multiput(-2.64,59.02)(0.06,-0.49){1}{\line(0,-1){0.49}}
\multiput(-2.58,58.53)(0.07,-0.49){1}{\line(0,-1){0.49}}
\multiput(-2.51,58.03)(0.08,-0.49){1}{\line(0,-1){0.49}}
\multiput(-2.43,57.54)(0.09,-0.49){1}{\line(0,-1){0.49}}
\multiput(-2.34,57.05)(0.1,-0.49){1}{\line(0,-1){0.49}}
\multiput(-2.24,56.57)(0.11,-0.48){1}{\line(0,-1){0.48}}
\multiput(-2.13,56.08)(0.12,-0.48){1}{\line(0,-1){0.48}}
\multiput(-2,55.6)(0.14,-0.48){1}{\line(0,-1){0.48}}
\multiput(-1.87,55.13)(0.15,-0.47){1}{\line(0,-1){0.47}}
\multiput(-1.72,54.65)(0.16,-0.47){1}{\line(0,-1){0.47}}
\multiput(-1.56,54.18)(0.17,-0.47){1}{\line(0,-1){0.47}}
\multiput(-1.39,53.71)(0.18,-0.46){1}{\line(0,-1){0.46}}
\multiput(-1.21,53.25)(0.1,-0.23){2}{\line(0,-1){0.23}}
\multiput(-1.02,52.79)(0.1,-0.23){2}{\line(0,-1){0.23}}
\multiput(-0.82,52.34)(0.11,-0.22){2}{\line(0,-1){0.22}}
\multiput(-0.61,51.89)(0.11,-0.22){2}{\line(0,-1){0.22}}
\multiput(-0.39,51.44)(0.12,-0.22){2}{\line(0,-1){0.22}}
\multiput(-0.16,51)(0.12,-0.22){2}{\line(0,-1){0.22}}
\multiput(0.08,50.57)(0.13,-0.21){2}{\line(0,-1){0.21}}
\multiput(0.34,50.14)(0.13,-0.21){2}{\line(0,-1){0.21}}
\multiput(0.6,49.72)(0.14,-0.21){2}{\line(0,-1){0.21}}
\multiput(0.87,49.3)(0.14,-0.2){2}{\line(0,-1){0.2}}
\multiput(1.15,48.89)(0.15,-0.2){2}{\line(0,-1){0.2}}
\multiput(1.44,48.49)(0.15,-0.2){2}{\line(0,-1){0.2}}
\multiput(1.74,48.09)(0.1,-0.13){3}{\line(0,-1){0.13}}
\multiput(2.05,47.7)(0.11,-0.13){3}{\line(0,-1){0.13}}
\multiput(2.36,47.32)(0.11,-0.12){3}{\line(0,-1){0.12}}
\multiput(2.69,46.95)(0.11,-0.12){3}{\line(0,-1){0.12}}
\multiput(3.03,46.58)(0.11,-0.12){3}{\line(0,-1){0.12}}
\multiput(3.37,46.22)(0.12,-0.12){3}{\line(0,-1){0.12}}
\multiput(3.72,45.87)(0.12,-0.11){3}{\line(1,0){0.12}}
\multiput(4.08,45.53)(0.12,-0.11){3}{\line(1,0){0.12}}
\multiput(4.45,45.19)(0.12,-0.11){3}{\line(1,0){0.12}}
\multiput(4.82,44.86)(0.13,-0.11){3}{\line(1,0){0.13}}
\multiput(5.2,44.55)(0.13,-0.1){3}{\line(1,0){0.13}}
\multiput(5.59,44.24)(0.2,-0.15){2}{\line(1,0){0.2}}
\multiput(5.99,43.94)(0.2,-0.15){2}{\line(1,0){0.2}}
\multiput(6.39,43.65)(0.2,-0.14){2}{\line(1,0){0.2}}
\multiput(6.8,43.37)(0.21,-0.14){2}{\line(1,0){0.21}}
\multiput(7.22,43.1)(0.21,-0.13){2}{\line(1,0){0.21}}
\multiput(7.64,42.84)(0.21,-0.13){2}{\line(1,0){0.21}}
\multiput(8.07,42.58)(0.22,-0.12){2}{\line(1,0){0.22}}
\multiput(8.5,42.34)(0.22,-0.12){2}{\line(1,0){0.22}}
\multiput(8.94,42.11)(0.22,-0.11){2}{\line(1,0){0.22}}
\multiput(9.39,41.89)(0.22,-0.11){2}{\line(1,0){0.22}}
\multiput(9.84,41.68)(0.23,-0.1){2}{\line(1,0){0.23}}
\multiput(10.29,41.48)(0.23,-0.1){2}{\line(1,0){0.23}}
\multiput(10.75,41.29)(0.46,-0.18){1}{\line(1,0){0.46}}
\multiput(11.21,41.11)(0.47,-0.17){1}{\line(1,0){0.47}}
\multiput(11.68,40.94)(0.47,-0.16){1}{\line(1,0){0.47}}
\multiput(12.15,40.78)(0.47,-0.15){1}{\line(1,0){0.47}}
\multiput(12.63,40.63)(0.48,-0.14){1}{\line(1,0){0.48}}
\multiput(13.1,40.5)(0.48,-0.12){1}{\line(1,0){0.48}}
\multiput(13.58,40.37)(0.48,-0.11){1}{\line(1,0){0.48}}
\multiput(14.07,40.26)(0.49,-0.1){1}{\line(1,0){0.49}}
\multiput(14.55,40.16)(0.49,-0.09){1}{\line(1,0){0.49}}
\multiput(15.04,40.07)(0.49,-0.08){1}{\line(1,0){0.49}}
\multiput(15.53,39.99)(0.49,-0.07){1}{\line(1,0){0.49}}
\multiput(16.03,39.92)(0.49,-0.06){1}{\line(1,0){0.49}}
\multiput(16.52,39.86)(0.49,-0.05){1}{\line(1,0){0.49}}
\multiput(17.01,39.81)(0.5,-0.03){1}{\line(1,0){0.5}}
\multiput(17.51,39.78)(0.5,-0.02){1}{\line(1,0){0.5}}
\multiput(18,39.76)(0.5,-0.01){1}{\line(1,0){0.5}}
\put(18.5,39.75){\line(1,0){0.5}}
\multiput(19,39.75)(0.5,0.01){1}{\line(1,0){0.5}}
\multiput(19.5,39.76)(0.5,0.02){1}{\line(1,0){0.5}}
\multiput(19.99,39.78)(0.5,0.03){1}{\line(1,0){0.5}}
\multiput(20.49,39.81)(0.49,0.05){1}{\line(1,0){0.49}}
\multiput(20.98,39.86)(0.49,0.06){1}{\line(1,0){0.49}}
\multiput(21.47,39.92)(0.49,0.07){1}{\line(1,0){0.49}}
\multiput(21.97,39.99)(0.49,0.08){1}{\line(1,0){0.49}}
\multiput(22.46,40.07)(0.49,0.09){1}{\line(1,0){0.49}}
\multiput(22.95,40.16)(0.49,0.1){1}{\line(1,0){0.49}}
\multiput(23.43,40.26)(0.48,0.11){1}{\line(1,0){0.48}}
\multiput(23.92,40.37)(0.48,0.12){1}{\line(1,0){0.48}}
\multiput(24.4,40.5)(0.48,0.14){1}{\line(1,0){0.48}}
\multiput(24.87,40.63)(0.47,0.15){1}{\line(1,0){0.47}}
\multiput(25.35,40.78)(0.47,0.16){1}{\line(1,0){0.47}}
\multiput(25.82,40.94)(0.47,0.17){1}{\line(1,0){0.47}}
\multiput(26.29,41.11)(0.46,0.18){1}{\line(1,0){0.46}}
\multiput(26.75,41.29)(0.23,0.1){2}{\line(1,0){0.23}}
\multiput(27.21,41.48)(0.23,0.1){2}{\line(1,0){0.23}}
\multiput(27.66,41.68)(0.22,0.11){2}{\line(1,0){0.22}}
\multiput(28.11,41.89)(0.22,0.11){2}{\line(1,0){0.22}}
\multiput(28.56,42.11)(0.22,0.12){2}{\line(1,0){0.22}}
\multiput(29,42.34)(0.22,0.12){2}{\line(1,0){0.22}}
\multiput(29.43,42.58)(0.21,0.13){2}{\line(1,0){0.21}}
\multiput(29.86,42.84)(0.21,0.13){2}{\line(1,0){0.21}}
\multiput(30.28,43.1)(0.21,0.14){2}{\line(1,0){0.21}}
\multiput(30.7,43.37)(0.2,0.14){2}{\line(1,0){0.2}}
\multiput(31.11,43.65)(0.2,0.15){2}{\line(1,0){0.2}}
\multiput(31.51,43.94)(0.2,0.15){2}{\line(1,0){0.2}}
\multiput(31.91,44.24)(0.13,0.1){3}{\line(1,0){0.13}}
\multiput(32.3,44.55)(0.13,0.11){3}{\line(1,0){0.13}}
\multiput(32.68,44.86)(0.12,0.11){3}{\line(1,0){0.12}}
\multiput(33.05,45.19)(0.12,0.11){3}{\line(1,0){0.12}}
\multiput(33.42,45.53)(0.12,0.11){3}{\line(1,0){0.12}}
\multiput(33.78,45.87)(0.12,0.12){3}{\line(1,0){0.12}}
\multiput(34.13,46.22)(0.11,0.12){3}{\line(0,1){0.12}}
\multiput(34.47,46.58)(0.11,0.12){3}{\line(0,1){0.12}}
\multiput(34.81,46.95)(0.11,0.12){3}{\line(0,1){0.12}}
\multiput(35.14,47.32)(0.11,0.13){3}{\line(0,1){0.13}}
\multiput(35.45,47.7)(0.1,0.13){3}{\line(0,1){0.13}}
\multiput(35.76,48.09)(0.15,0.2){2}{\line(0,1){0.2}}
\multiput(36.06,48.49)(0.15,0.2){2}{\line(0,1){0.2}}
\multiput(36.35,48.89)(0.14,0.2){2}{\line(0,1){0.2}}
\multiput(36.63,49.3)(0.14,0.21){2}{\line(0,1){0.21}}
\multiput(36.9,49.72)(0.13,0.21){2}{\line(0,1){0.21}}
\multiput(37.16,50.14)(0.13,0.21){2}{\line(0,1){0.21}}
\multiput(37.42,50.57)(0.12,0.22){2}{\line(0,1){0.22}}
\multiput(37.66,51)(0.12,0.22){2}{\line(0,1){0.22}}
\multiput(37.89,51.44)(0.11,0.22){2}{\line(0,1){0.22}}
\multiput(38.11,51.89)(0.11,0.22){2}{\line(0,1){0.22}}
\multiput(38.32,52.34)(0.1,0.23){2}{\line(0,1){0.23}}
\multiput(38.52,52.79)(0.1,0.23){2}{\line(0,1){0.23}}
\multiput(38.71,53.25)(0.18,0.46){1}{\line(0,1){0.46}}
\multiput(38.89,53.71)(0.17,0.47){1}{\line(0,1){0.47}}
\multiput(39.06,54.18)(0.16,0.47){1}{\line(0,1){0.47}}
\multiput(39.22,54.65)(0.15,0.47){1}{\line(0,1){0.47}}
\multiput(39.37,55.13)(0.14,0.48){1}{\line(0,1){0.48}}
\multiput(39.5,55.6)(0.12,0.48){1}{\line(0,1){0.48}}
\multiput(39.63,56.08)(0.11,0.48){1}{\line(0,1){0.48}}
\multiput(39.74,56.57)(0.1,0.49){1}{\line(0,1){0.49}}
\multiput(39.84,57.05)(0.09,0.49){1}{\line(0,1){0.49}}
\multiput(39.93,57.54)(0.08,0.49){1}{\line(0,1){0.49}}
\multiput(40.01,58.03)(0.07,0.49){1}{\line(0,1){0.49}}
\multiput(40.08,58.53)(0.06,0.49){1}{\line(0,1){0.49}}
\multiput(40.14,59.02)(0.05,0.49){1}{\line(0,1){0.49}}
\multiput(40.19,59.51)(0.03,0.5){1}{\line(0,1){0.5}}
\multiput(40.22,60.01)(0.02,0.5){1}{\line(0,1){0.5}}
\multiput(40.24,60.5)(0.01,0.5){1}{\line(0,1){0.5}}

\linethickness{0.2mm}
\qbezier(13.25,81.68)(13.11,81.6)(12.25,80.58)
\qbezier(12.25,80.58)(11.4,79.55)(11.49,78.61)
\qbezier(11.49,78.61)(11.86,77.87)(13.03,77.74)
\qbezier(13.03,77.74)(14.2,77.61)(14.56,76.86)
\qbezier(14.56,76.86)(14.63,76.03)(13.68,75.33)
\qbezier(13.68,75.33)(12.74,74.62)(12.81,73.79)
\qbezier(12.81,73.79)(13.17,73.04)(14.34,72.91)
\qbezier(14.34,72.91)(15.51,72.79)(15.88,72.04)
\qbezier(15.88,72.04)(15.94,71.21)(15,70.5)
\qbezier(15,70.5)(14.06,69.8)(14.12,68.97)
\qbezier(14.12,68.97)(14.49,68.22)(15.66,68.09)
\qbezier(15.66,68.09)(16.83,67.96)(17.19,67.21)
\qbezier(17.19,67.21)(17.26,66.38)(16.32,65.68)
\qbezier(16.32,65.68)(15.37,64.97)(15.44,64.14)
\qbezier(15.44,64.14)(15.8,63.39)(16.97,63.27)
\qbezier(16.97,63.27)(18.14,63.14)(18.51,62.39)
\linethickness{0.2mm} \qbezier(-2.5,61.25)(-2.45,61.09)(-1.69,60)
\qbezier(-1.69,60)(-0.93,58.91)(0,58.75)
\qbezier(0,58.75)(0.82,58.91)(1.25,60)
\qbezier(1.25,60)(1.68,61.09)(2.5,61.25)
\qbezier(2.5,61.25)(3.32,61.09)(3.75,60)
\qbezier(3.75,60)(4.18,58.91)(5,58.75)
\qbezier(5,58.75)(5.82,58.91)(6.25,60)
\qbezier(6.25,60)(6.68,61.09)(7.5,61.25)
\qbezier(7.5,61.25)(8.32,61.09)(8.75,60)
\qbezier(8.75,60)(9.18,58.91)(10,58.75)
\qbezier(10,58.75)(10.82,58.91)(11.25,60)
\qbezier(11.25,60)(11.68,61.09)(12.5,61.25)
\qbezier(12.5,61.25)(13.32,61.09)(13.75,60)
\qbezier(13.75,60)(14.18,58.91)(15,58.75)
\qbezier(15,58.75)(15.82,58.91)(16.25,60)
\qbezier(16.25,60)(16.68,61.09)(17.5,61.25) \linethickness{0.2mm}
\qbezier(38.89,69.27)(38.79,69.4)(37.68,70.14)
\qbezier(37.68,70.14)(36.57,70.88)(35.65,70.69)
\qbezier(35.65,70.69)(34.94,70.24)(34.94,69.07)
\qbezier(34.94,69.07)(34.94,67.89)(34.23,67.45)
\qbezier(34.23,67.45)(33.41,67.3)(32.61,68.16)
\qbezier(32.61,68.16)(31.81,69.02)(30.99,68.87)
\qbezier(30.99,68.87)(30.29,68.42)(30.28,67.25)
\qbezier(30.28,67.25)(30.28,66.07)(29.58,65.63)
\qbezier(29.58,65.63)(28.76,65.47)(27.96,66.34)
\qbezier(27.96,66.34)(27.16,67.2)(26.34,67.04)
\qbezier(26.34,67.04)(25.63,66.6)(25.63,65.42)
\qbezier(25.63,65.42)(25.62,64.25)(24.92,63.81)
\qbezier(24.92,63.81)(24.1,63.65)(23.3,64.51)
\qbezier(23.3,64.51)(22.5,65.38)(21.68,65.22)
\qbezier(21.68,65.22)(20.97,64.78)(20.97,63.6)
\qbezier(20.97,63.6)(20.97,62.43)(20.26,61.98)
\linethickness{0.2mm} \qbezier(4.09,45.87)(4.24,45.79)(5.55,45.59)
\qbezier(5.55,45.59)(6.87,45.39)(7.63,45.95)
\qbezier(7.63,45.95)(8.08,46.65)(7.58,47.72)
\qbezier(7.58,47.72)(7.09,48.79)(7.54,49.49)
\qbezier(7.54,49.49)(8.22,49.97)(9.31,49.53)
\qbezier(9.31,49.53)(10.4,49.09)(11.07,49.57)
\qbezier(11.07,49.57)(11.53,50.27)(11.03,51.34)
\qbezier(11.03,51.34)(10.54,52.41)(10.99,53.11)
\qbezier(10.99,53.11)(11.67,53.59)(12.76,53.15)
\qbezier(12.76,53.15)(13.85,52.71)(14.52,53.19)
\qbezier(14.52,53.19)(14.98,53.89)(14.48,54.96)
\qbezier(14.48,54.96)(13.99,56.03)(14.44,56.73)
\qbezier(14.44,56.73)(15.12,57.21)(16.21,56.77)
\qbezier(16.21,56.77)(17.3,56.33)(17.97,56.81)
\qbezier(17.97,56.81)(18.43,57.51)(17.93,58.58)
\qbezier(17.93,58.58)(17.44,59.65)(17.89,60.35)
\linethickness{0.2mm}
\qbezier(34.29,46.86)(34.36,47.01)(34.5,48.34)
\qbezier(34.5,48.34)(34.65,49.66)(34.05,50.39)
\qbezier(34.05,50.39)(33.33,50.81)(32.28,50.27)
\qbezier(32.28,50.27)(31.24,49.72)(30.52,50.14)
\qbezier(30.52,50.14)(30,50.8)(30.4,51.91)
\qbezier(30.4,51.91)(30.79,53.02)(30.27,53.67)
\qbezier(30.27,53.67)(29.55,54.09)(28.51,53.55)
\qbezier(28.51,53.55)(27.47,53)(26.75,53.42)
\qbezier(26.75,53.42)(26.23,54.08)(26.62,55.19)
\qbezier(26.62,55.19)(27.01,56.3)(26.5,56.95)
\qbezier(26.5,56.95)(25.78,57.37)(24.74,56.83)
\qbezier(24.74,56.83)(23.69,56.28)(22.97,56.7)
\qbezier(22.97,56.7)(22.46,57.36)(22.85,58.47)
\qbezier(22.85,58.47)(23.24,59.58)(22.73,60.23)
\qbezier(22.73,60.23)(22,60.65)(20.96,60.11)
\qbezier(20.96,60.11)(19.92,59.56)(19.2,59.98)
\linethickness{0.2mm} \put(18.75,61.25){\circle{3.54}}
\end{picture}
}

\newcommand{\aei}{\it Max Planck Institute for Gravitational Physics
(Albert Einstein Institute)\\ Am M\"uhlenberg 1, 14476 Golm,
Germany}

\newcommand{\lpi}{\it Lebedev Institute of Physics, Leninskiy prospekt, 53, 119991, Moscow, Russia}
\begin{document}
\thispagestyle{empty}
\begin{flushright}
\hfill{AEI-2013-034}
\end{flushright}
\begin{center}

~\vspace{20pt}

{\Large\bf Exact higher-spin symmetry in CFT: \\ free fermion correlators from Vasiliev Theory}

\vspace{25pt}

V.E. Didenko\symbolfootnote[1]{Email:~\sf didenko@lpi.ru}, Jianwei
Mei\symbolfootnote[2]{Email:~\sf jwmei@aei.mpg.de} and E.D.
Skvortsov\symbolfootnote[3]{Email:~\sf skvortsov@lpi.ru}

\vspace{10pt}${}^\ast{}^\ddagger$\lpi

\vspace{10pt}${}^\dagger{}^\ddagger$\aei

\vspace{2cm}

\begin{abstract}
N-point correlation functions of conserved currents and
weight-two scalar operators of the three-dimensional free fermion
vector model are found as invariants of the higher-spin symmetry
in four-dimensional AdS. These are the correlators of the unbroken
Vasiliev higher-spin theory. The results extend the recent work
arXiv:1210.7963 and are complementary to arXiv:1301.3123 where the
correlators were computed entirely on the boundary.
\end{abstract}

\end{center}

 \newpage

\tableofcontents

\section{Introduction}
Higher-spin theories constructed by Vasiliev
\cite{Vasiliev:1990en, Vasiliev:1990vu, Vasiliev:1992av,
Vasiliev:1995dn, Vasiliev:1999ba, Vasiliev:2003ev} have attracted
much attention \cite{Giombi:2009wh, Giombi:2010vg,
Henneaux:2010xg, Campoleoni:2010zq, Colombo:2012jx,
Didenko:2012vh, Vasiliev:2012vf, Chang:2012kt} as simple models of
AdS/CFT, \cite{Maldacena:1997re, Gubser:1998bc, Witten:1998qj},
with the CFT duals being vector models, some of which are of
practical value, e.g., the critical $O(N)$ model,
\cite{Sezgin:2002rt, Klebanov:2002ja,Sezgin:2003pt}.

The higher spin (HS) symmetry is an infinite-dimensional extension
of the conformal symmetry and is strong enough to fix the form of
all correlations functions, \cite{Maldacena:2011jn}. However, the
Maldacena-Zhiboedov theorem \cite{Maldacena:2011jn}, which can be
thought of as an extension of the Coleman-Mandula no-go theorem,
tells us that under some mild assumptions such as unitarity,
locality\footnote{Let us note that in context of HS theories the
notion of local CFTs should be treated with great care as the bulk
theories are nonlocal.} and OPE it is impossible to have
interacting CFTs that allow infinitely many conserved HS charges.
In other words, if the bulk theory admits such a boundary behavior
that leaves HS symmetry unbroken, the corresponding dual CFT is
free. On the other hand, the extension of the Maldacena-Zhiboedov
result \cite{Maldacena:2012sf} shows that broken HS symmetry still
restricts correlations functions and thus can be effectively used
as a source of nontrivial integrable models once the mechanism of
breaking is understood. As a starting point it is required to
understand how the exact HS symmetry can be used to efficiently
determine the form of correlations functions.

Recently by using the HS symmetry as a higher-dimensional
replacement of the Virasoro algebra, the correlation functions of
all orders of conserved currents in the three-dimensional CFT's
that have exact HS symmetry have been found in
\cite{Didenko:2012tv}. By the Maldacena-Zhiboedov theorem these
are free theories, either free boson or free fermion. The main
goal of \cite{Didenko:2012tv} was to give an explicit formula for
all correlators relying on symmetry requirements only.

The constructive formula for $n$-point correlation function
proposed in \cite{Didenko:2012tv} reads
\be \langle j(x_1,\eta_1)\cdots j(x_n,\eta_n)\rangle
=\sum_{S_n}\mbox{Tr}(\Psi(x,x_1,\eta_1)\star\cdots\star
\Psi(x,x_n,\eta_n))\,,\label{allcorrfunc} \ee
and it is analogous to the definition of long-trace operators,
with the difference being that the trace is taken in the
infinite-dimensional HS algebra rather than in $SU(N)$. Let us now
explain the constituents of (\ref{allcorrfunc}) in detail. On the
CFT side we focus on symmetric and traceless tensors,
$j_{a_1\cdots a_s}$, which are in addition conserved $\partial^m
j_{m a_2...a_s}=0$ and thus are primary fields to be referred to
as currents. It is convenient to pack all currents $j_{a_1\cdots
a_s}$ into a generating function
\be j(x,\eta)=\sum_s j_{a_1\cdots a_{s}}\,\eta^{a_1} \cdots
\eta^{a_s}=\sum_s j_{\alpha_1\cdots
\alpha_{2s}}\,\eta^{\alpha}\cdots\eta^{\alpha}\,,\ee
where $\eta^a$'s are null polarization vectors, $\eta^a\eta_a=0$.
In $3d$ the spinor language has a great advantage and instead of
light-like vector $\eta^a$ we introduced  two-component spinor
$\eta^\alpha$. Then, given the {\em l.h.s.} of
\eqref{allcorrfunc}, one can extract the $n$-point correlation
function of some particular currents of spins $s_1,\cdots s_n$ as
the order $2s_1,\cdots,2s_n$ Taylor coefficient of
$\eta^\alpha_1,\cdots,\eta^\alpha_n$. The definition of the {\em
r.h.s.} of \eqref{allcorrfunc} requires a certain AdS/CFT inspired
technique that was laid down in
\cite{Sezgin:2005pv,Colombo:2010fu,Colombo:2012jx}. The key object
$\Psi(x,x_i,\eta_i)$ is the Fourier transform with respect to some
auxiliary variable of the boundary-to-bulk propagator for the
master field-strength of the Vasiliev HS theory in $AdS_4$. It
depends on the $AdS_4$ coordinate $x$; on the boundary point $x_i$
where the current $j(x_i,\eta_i)$ is inserted; on the polarization
$\eta_i^\alpha$ that encodes the index structure; on the auxiliary
variable $Y$ that generates HS algebra and was left implicit.

There are a few important facts about $\Psi(x,x_i,\eta_i)$.
Firstly, it behaves like a (set of) conserved current with respect
to $x_i$ and $\eta_i$. Secondly, it transforms in the adjoint of
the HS algebra
\be\delta\Psi=[\Psi,\xi]_\star\,,\label{trans.Phi}\ee
where $\star$ denotes the product with respect to $Y$ in the HS
algebra. It is also assumed that the HS algebra admits a trace,
which has the right property to make \eqref{allcorrfunc} invariant
under all HS transformations (\ref{trans.Phi}). In particular,
since the conformal algebra is itself a subalgebra of the HS
algebra, \eqref{allcorrfunc} is conformally-invariant and behaves
as a conserved current in each slot. From the bulk point of view,
a large adjoint transformation $\Psi\rightarrow
g^{-1}(x)\star\Psi\star g(x)$, where $g(x)$ occupies only the
$SO(3,2)$ subgroup, allows one to move $x$ freely in the $AdS$.
Therefore, the dependence on the bulk point $x$ drops out of
\eqref{allcorrfunc}. Lastly, the sum over the symmetric group
makes the result symmetric in its arguments. From the bulk point
of view the sum is  necessary to have the trace
\eqref{allcorrfunc} real, that is the symmetrization is driven by
appropriate reality conditions for master fields.

\begin{wrapfigure}{l}{0.3\textwidth}
\vspace{5pt}
\begin{center}
$$\TwistorialWitten$$
\end{center}
\end{wrapfigure}
One can also view \eqref{allcorrfunc} as originating from a
Witten-like diagram (left) for a vertex \be\label{tracecyc}
V_n=\mbox{Tr}(\Psi\star...\star \Psi)\,.\ee Since the trace does
not depend on the interaction point $x$ in the bulk the integral
over $AdS$ drops out. The sum of such diagrams seems to be what
the Vasiliev theory reduces to with the boundary conditions that do
not break HS symmetry.

Let us note that the formula (\ref{allcorrfunc}) is quite general
and can be applied to any CFT with HS symmetry, e.g. free scalar
and boson in $d$-dimensions, free limit of SYM, generalized free
fields and perhaps to the duals of $3d$ HS theories, etc.

According to \cite{Klebanov:2002ja} and \cite{Sezgin:2003pt} the
currents $j_{a_1\cdots a_s}$ should be originated either from free
scalar $\phi$ or from free fermion $\psi$. These are the conformal
primaries appearing in the OPE $\phi\times\phi$ or
$\psi\times\psi$. The only difference between free boson and free
fermion at the level of currents $j_{a_1\cdots a_s}$ is that the
first member of the family $j_0$, $j_0=\phi^2$ or $j_0=\psi^2$,
which is not a genuine current, may have different conformal
weights, $\Delta=1$ or $\Delta=2$ depending on wether it is made
of a boson or a fermion, respectively. In \cite{Didenko:2012tv}
only the case with $\Delta=s+1$ boundary conditions was
considered, while the case of $\Delta=2$ operator, which we will
refer to as $\td{j}_0$, was not included. Therefore, the part of
the correlators of the free fermion model were not reproduced.
This is the gap we would like fill in the present paper, so we
would like to compute various correlators of the form
\be\label{corr} \brk{\td{j}_0\cdots\td{j}_0\, j(x_1,\eta_1)\cdots
j(x_n,\eta_n)}\,. \ee

The results of the paper shows that \eqref{allcorrfunc} works
well, giving all correlators of the operators that are dual to the
higher-spin multiplet in the bulk of $AdS_4$. Our results are
closely related to the recent paper \cite{Gelfond:2013xt} by
Gelfond and Vasiliev. In \cite{Gelfond:2013xt} the operator
product algebra of free boson and free fermion was found
explicitly and then used to compute all the correlators. The
advantage of having the operator algebra at hand is that the
overall prefactors, which are left undetermined in our approach,
can be fixed in terms of $N$ ($N$ is the number of free fields in
the multiplet). Complementary to \cite{Gelfond:2013xt} our method
relies on the AdS/CFT and provides a link between the computations
entirely in the bulk and on the boundary. We expect that our basic
formula \eqref{allcorrfunc} is the prolongation of
\cite{Gelfond:2013xt} to the bulk of $AdS$. At the same time our
work and \cite{Didenko:2012tv} are similar to
\cite{Gelfond:2013xt} in a sense that both approaches are in fact
$sp(2M)$ covariant with $M=2$ for the case of interest
$AdS_4/CFT_3$ since $so(3,2)\sim sp(4)$. Particularly, using this
fact the authors of \cite{Gelfond:2013xt} were able to
straightforwardly generalize their results to reproduce
correlation functions of $4d$ conserved currents as being realized
via embedding into $sp(8)$. These formally coincide\footnote{We
mean that the dependence on the tensor conformally-invariant
structures is the same, but the prefactors $|x_i-x_j|^{-\delta}$
come with powers $\delta$ that depend on $d$ of course.} with
those of \cite{Didenko:2012tv} as the basic building blocks, the
conformally-invariant structures, remain the same as well as the
general $sp(2M)$ formula of \cite{Didenko:2012tv} used to derive
the correlation functions.

The outline of the paper is as follows. For reader's convenience
we summarize our key results in the next section and then present
our derivation in the following sections. In particular, we derive
the generating functions and correlation functions in section 3.
The conclusions are in section \ref{sec:conclusions}. Certain
technicalities are collected in Appendix \ref{secDetails}.

In this paper we use the same strategy and techniques as developed
in \cite{Didenko:2012tv}, where all the results for $\Delta=s+1$
have been obtained. For reader's convenience, in our derivation we
will quote the main results of \cite{Didenko:2012tv} and put them
side by side with the new results obtained involving $\td{j}_0$.

\section{Main results: Examples}

In the Vasiliev higher-spin theory there is a free
parameter\footnote{There are infinitely many free constants in the
parity-violating Vasiliev theory, $\theta$ being the first of
them. We expect that the effect of the rest of the constants is
just to renormalize $\theta$. } $\theta$ that allows one to
interpolate smoothly between the duals. Our results correspond to
the free boson at $\theta=0$ and free fermion at $\theta=\pi/2$.
We prefer to keep $\theta$ everywhere. The correlators we consider
are connected correlators and they are defined up to an overall
factor, which cannot be fixed in \eqref{allcorrfunc}.

\begin{table}
\begin{center}
\begin{tabular}{|c|c|c|c|}
 \hline
 Conformal                    & Explicit coordinate                                          & Number    & Parity\\
 structures                   & representation                                               & of points &\\
 \hline
 \rule{0pt}{20pt}$P_{ij}$     & $\eta_i\bx_{ij}^{-1}\eta_j$                                  & 2         & even\\
 \rule{0pt}{20pt}$Q^{i}_{jk}$ & $\eta_i[\bx_{ij}^{-1}-\bx_{ik}^{-1}]\eta_i$                  & 3         & even\\
 \rule{0pt}{20pt}$S_{ik}^j$   & $\ds\frac{\eta_i\bx_{ij}\bx_{jk}\eta_k}{x_{ij}x_{jk}x_{ki}}$ & 3         & odd \\
 \rule{0pt}{20pt}$R_{ij}$     & $\eta_i\bx_{i(i+1)}^{-1}\cdots\bx_{(j-1)j}^{-1}\eta_j$
 & $j-i+1$ & $(-)^{j-i+1}$ \\
 \hline
\end{tabular}
\end{center}
\caption{\it Conformal structures. Note $R$ is not independent
from $P$, $Q$ or $S$. Coordinates on the boundary are
parameterized by symmetric bispinors $\bx\equiv (\bx^{\ga\gb})
=(\bx^{\gb\ga})$ and we have chosen to suppress the spinor
indices. The subscript on $\bx_i$ refers to the $i$-th point at
which some operator is inserted, $\eta_i$ is the boundary
polarization, $\bx_{ij}=\bx_i -\bx_j$ and
$x_{ij}=\sqrt{-\det\bx_{ij}}$ is the distance between the $i$'th
point and the $j$'th point. Parity is determined under the
transformation $\bx\to-\bx$ and $\eta\to i\eta$, with positive
being even and negative being odd.}\label{table.PQRS}
\end{table}

Before going to the examples, note for tensor operators it is
convenient to use conformally invariant structures, i.e.  $P$, $Q$
and $S$, as introduced in \cite{Costa:2011mg, Giombi:2011rz}.
However, we find that starting from two $\tilde{j}_0$-insertions
it is no longer convenient to make use of the $ S$ structure,
especially in correlators of orders higher than three. Instead, we
find it more convenient to introduce a new object $R$. Some basic
properties of these structures are listed in Table
\ref{table.PQRS}. In our calculation, we will need the following
(They only differ from those in Table \ref{table.PQRS} by
numerical factors),
\bea Q^i&=&-\frac18\eta_i[\bx_{i(i+1)}^{-1}
-\bx_{i(i-1)}^{-1}]\eta_i\,,\nn\\
P_{i(i+1)}&=&\frac14\eta_i\bx_{i(i+1)}^{-1}\eta_{i+1}\,,\quad
1\leq i\leq n-1\,,\nn\\
P_{01}&=&P_{n(n+1)}=P_{n1}=-\frac14\eta_n\bx_{n1}^{-1}\eta_1\,,\nn\\
R_{jk}&=&-\ds\frac{c}{2(2i)^{k-j}}\,\eta_j\bx_{j(j+1)}^{-1}
\cdots\bx_{(k-1)k}^{-1}\eta_k\,,\nn\\
R_{0j}&=&R_{nj}\,,\quad R_{j(n+1)}=R_{j0}\,,\quad\forall~j\,,\eea
where $n$ is the order of correlation functions. The meaning of the
above definitions is that $P$, $Q$, $S$ and $R$ structures involve
a pair or a triple of the points adjacent along the cycle $12...n$
for the $n$-point function. Note that
\be R_{jk}\sim\eta_j\bx_{j(j+1)}^{-1}\cdots \bx_{(n-1)n}^{-1}
\bx_{n1}^{-1}\cdots\bx_{(k-1)k}^{-1}\eta_k\ee
if $j>k$ (after the replacement $R_{0j}\to R_{nj}$ or
$R_{j(n+1)}\to R_{j0}$, if any). Finally, $c=1$ if $R_{jk}$ does
not contain $\bx_{n1}^{-1}$ and $c=-1$ otherwise.

In the examples below $j_s$ refers to the insertion of
$j_{a_1...a_s}$, i.e. $\Delta=s+1$ operator. In particular the
generating function corresponding to $j_s$ contains
$(\Delta=1)$-operator $j_0=\phi^2$.  The insertion of the
$(\Delta=2)$-operator $\psi^2$ is denoted by $\td{j}_0$. Of
course, all correlators involving $j_0$ and $\td{j}_0$
simultaneously must vanish.

\paragraph{\bf Two-point functions}
Two point functions are fixed by conformal symmetry up to a
number:
\bea\brk{j_sj_s}&\propto&\frac1{x_{12}^2}\Big(\cos^2\theta\cos^2
 P_{12}+\sin^2\theta\sin^2 P_{12}\Big)\,,\nn\\
\brk{\td{j}_0\td{j}_0}&\propto&\frac1{x_{12}^4}\,,
\label{exmp.2pt}\eea

\paragraph{Three-point functions}
Three point functions are known to be fixed up to a number of
constants:
\bea\brk{j_sj_sj_s}&\propto&\frac1{x_{12}x_{23}x_{31}}\Big[\cos^3\theta
\cos( Q^1+ Q^2+ Q^3)\cos P_{12}\cos P_{23}\cos P_{31}\nn\\
&&\qquad\qquad+i\sin^3\theta\sin( Q^1+ Q^2+ Q^3)\sin P_{12}
\sin P_{23}\sin P_{31}\Big]\,,\nn\\
\brk{\td{j}_0j_sj_s}&\propto&\sin^2\theta\frac{\cos( Q^2+
Q^3)}{x_{12}
x_{23}x_{31}} R_{02}\sin P_{23}\,,\nn\\
\brk{\td{j}_0\td{j}_0j_s}&\propto&\sin\theta\frac{\sin Q^3}{x_{12}
x_{23}x_{31}} R_{03}\,,\nn\\
\brk{\td{j}_0\td{j}_0\td{j}_0}&=&0\,.\label{exmp.3pt}\eea
The first formula, borrowed from \cite{Colombo:2012jx,
Didenko:2012tv}, correctly reproduces three-point functions of
free boson (first part) and three-point functions of the conserved
currents, $j_s$, $s\geq1$ of free fermion (second part). These two
contributions were previously obtained by solving Vasiliev
equations to the second order in \cite{Giombi:2010vg}. The second
formula coincides with the one from \cite{Giombi:2011rz}, which
was claimed to have been obtained from the Vasiliev theory too.
Let us note the appearance of the odd conformally-invariant
structure\footnote{$S$ as defined in \cite{Giombi:2011rz} contains
a factor of $P$, which we removed from our definition.} $S$, which
is a particular case of our $ R$-sructure: $R_{02}\propto\eta_3
\bx_{31}^{-1} \bx_{12}^{-1} \eta_2\sim S^1$. The third correlator
is in fact fixed up to an overall coefficient by the conformal
symmetry and in this degenerate case the $ R$-structure coincides
with $Q$: $ R_{03}\propto\eta_3 \bx_{31}^{-1} \bx_{12}^{-1}
\bx_{23}^{-1} \eta_3\sim Q^3$. The last correlator, which vanishes
identically irrespectively of $\theta$, is correct both for the
free fermion theory and the critical boson. The latter seems to be
accidental as our considerations based on the exact HS symmetry do
not apply to the critical boson. All the three-point functions
above can be found also in \cite{Gelfond:2013xt}.

\paragraph{Four-point functions}
The four-point functions were missing in the literature before
\cite{Didenko:2012tv} apart from a very specific correlators used
in \cite{Maldacena:2011jn},
\bea\brk{j_sj_sj_sj_s}&\propto&\sum_{\text{perm}}\frac{\cos(Q^1
+Q^2+Q^3+Q^4)}{x_{12}x_{23} x_{34}x_{31}}\Big(\cos^4\theta\cos
P_{12}\cos P_{23}\cos P_{34}\cos P_{41}\nn\\
&&\qquad\qquad\qquad\qquad\qquad\qquad+\sin^4\theta\sin P_{12}
\sin P_{23}\sin P_{34}\sin P_{41}\Big)\,,\nn\\
\brk{\td{j}_0j_sj_sj_s}&\propto&\sin^3\theta\sum_{\text{perm}}
\frac{\sin(Q^2+Q^3+Q^4)}{x_{12}x_{23}x_{34}x_{31}} R_{02}\sin
P_{23} \sin P_{34}\,,\nn\\
\brk{\td{j}_0\td{j}_0j_sj_s}&\propto&\sin^2\theta\sum_{\text{perm}}\Big[\frac{\cos(Q^2+Q^4)}{x_{12}x_{23}x_{34}x_{41}}R_{02}R_{24}\Big|_{\xi_2\to\xi_3,\bx_2\leftrightarrow\bx_3}+\frac{\cos(Q^3+Q^4)}{x_{12}x_{23}x_{34}x_{41}}iR_{03}
\sin P_{34}\Big]
\,,\nn\\
\brk{\td{j}_0\td{j}_0\td{j}_0j_s}&\propto&\sin\theta\sum_{\text{perm}}\frac{\sin Q^4}{x_{12}x_{23}x_{34}x_{31}}R_{04}\,,\nn\\
\brk{\td{j}_0\td{j}_0\td{j}_0\td{j}_0}&\propto&\sum_{\text{perm}}\frac{tr(\bx_{ 12}^{-1}\bx_{23}^{-1}\bx_{34}^{-1} \bx_{
41}^{-1})}{x_{12}x_{23} x_{34}x_{41}}\,,\label{exmp.4pt}\eea
where $\sum_\text{perm}$ implies the sum over the necessary permutations of the legs. The first formula is borrowed from \cite{Didenko:2012tv}. Despite
containing two-pieces, first of the free boson and second of the
free fermion, the free fermion piece vanishes when at least one of
the polarizations $\eta$ is set to zero in accordance with the
fact that $\Delta=s+1$ propagator cannot account for the
$\Delta=2$ operator $\tilde{j}_0$. The correlators with different
number of insertions of $\tilde{j}_0$ are given afterwards. In
particular, $\brk{\td{j}_0\td{j}_0\td{j}_0\td{j}_0}$ depends
nontrivially on the two conformally-invariant ratios and this
dependence is reproduced by the enumerator. It is worth mentioning that the correlation functions consist of several independent parts whenever there are several ways to distribute $\td{j}_0$ in between $j_s$'s. Such parts are separately higher spin invariant and hence are separately conserved in each of the $j_s$. For example, the two terms in $\brk{\td{j}_0\td{j}_0j_sj_s}$ are conserved separately. 

The general case of $n$-point correlation functions requires more
technicalities and is discussed below.

\section{Generating functions and correlation functions}

In this section, we will calculate (\ref{allcorrfunc}) specified
to the $4d$ Vasiliev theory. A few basic definitions are given
below:
\begin{itemize}
\item The $4d$ HS algebra is the Weyl algebra with $sp(4)$ vectors
$Y^A$ as generating elements obeying
\be[Y_A,\,Y_B]=2i\epsilon_{AB}\,,\ee
where $\epsilon_{AB}$ is the $sp(4)$ invariant metric, which is
used to raise and lower indices $Y^A=\epsilon^{AB}Y_B$,
$Y_A=Y^B\epsilon_{BA}$. It is convenient to use the Weyl
$\star$-product realization. Then, the elements of the HS algebra
are functions of formally commuting variables $Y_A$ with the
product
\bea\label{star} f(Y)\star g(Y)=\int\, dU\,dV\,
f(Y+U)g(Y+V)e^{iV^{A} U_{A}}=f(Y)\exp
\left\{i\overleftarrow{\partial}_A\epsilon^{AB}
\overrightarrow{\partial}_B\right\}g(Y)\,.\eea
In the calculation, we will suppress spinor indices in most
places.\footnote{For this purpose, we follow the convention that
implicit spinor indices will always be contracted from the
upper-left to the lower-right direction, e.g. $YMY\equiv
Y^AM_A^{~B}Y_B$. To be consistent with this, the implicit index
positions on a matrix are always $M=M_\bullet^{~\bullet}$ and a
generalized notion of ``transpose", $\breve{M}_\bullet^{~\circ}
=\epsilon^{\circ\diamond}(M^T)^\ast_{ ~\diamond}\epsilon_{\ast
\bullet}$, is introduced so that indices on a transposed matrix
are also at the correct positions. Note the relation
$\epsilon^{AB}=\mbox{diag}(\epsilon^{\alpha\beta},
\epsilon^{\dot\alpha\dot\beta})$.}

\item The adjoint HS field $\Psi$ in (\ref{allcorrfunc}) is a
Fourier transform of the boundary-to-bulk propagator for the
master field-strength $B$, \cite{Vasiliev:1990en, Vasiliev:1990vu,
Vasiliev:1992av, Vasiliev:1995dn, Vasiliev:1999ba}. Namely,
$\Psi=B\star\delta$ and $\delta(y)=\int\frac{dp}{2\pi}e^{ipy}$.
Here $Y^A=(y^\alpha, \bar{y}^{\dot{\alpha}})$, i.e.
$Y=(y,\bar{y})$.

\item We need the boundary-to-bulk propagator of $B$ with
$\Delta=s+1$ boundary conditions, which describes bosonic fields
of all spins in the bulk, and $\Delta=2$ propagator for the scalar
component of the HS multiplet. The propagators,
\cite{Giombi:2009wh, Giombi:2010vg}, see also
\cite{Didenko:2012tv}, read
\bea \Delta=s+1\qquad B_i=B(x,x_i,\eta_i)&=&K_ie^{-iyF_i\bar{y}}
\Big(e^{-iy\xi_i +i\theta}+e^{iy \xi_i+i\theta}+e^{-i\bar{y}
\bar\xi_i-i\theta}+e^{i\bar{y}\bar\xi_i-i\theta}\Big)\,,
\label{def.B1.boson}\\
\Delta=2\qquad\qquad B'_i=B'(x,x_i)&=&K_i^2(1-iyF_i\bar{y})
e^{-iyF_i\bar{y}}\,,\label{def.B1prime}\eea
where the spinors $\xi$ and $\bar\xi$ are the bulk
polarization spinors, these are obtained by the parallel transport
of the boundary polarization $\eta_i$ to the bulk. $F\equiv
F^{\alpha\dot{\alpha}}$ is the wave vector from the bulk point $x$
towards the boundary point $x_i$. $K_i$ is the Witten $\Delta=1$
propagator for the scalar field. More details given in Poincare
coordinates can be found in Appendix
\ref{secDetails}.
\end{itemize}

\subsection{Strategy}

Given the above definitions, we will take the following steps to
calculate (\ref{allcorrfunc}):

{\bf 1.} Suppose that there are $n$ currents on the boundary, we
have
\be B=\sum_{i=1}^nB_i{\rm~or~}B'_i\,.\ee
Our strategy to compute $V_n=Tr((B\star\delta)^n)$ is to calculate
first $\cZ_n={\rm Tr}(B_1 \star \delta\star\cdots\star
B_n\star\delta)$ and then to apply the permutation group
$\mathbb{S}_n$ on $\cZ_n$ to obtain $V_n$.\footnote{We will not
consider contact terms in this calculation. So in fact we are
calculating $$V_n=Tr((B\star\delta)^n)- ({\rm contact
~terms})\,.$$}

{\bf 2.} To calculate $\cZ_n$, it will be convenient to start with
\be\Phi_i=K_ie^{-iyF_i\bar{y}-iy\xi_i-i\bar{y}\bar\xi_i+i\Theta_i}
=K_ie^{-\frac{i}2Yf_iY-iY\Xi_i+i\Theta_i}\,,\quad f_i=\left(
\begin{matrix}&F_i\cr\breve{F}_i&\end{matrix}\right)\,, \label{def.Phi}\ee
where $\Xi^A=\{\xi^\alpha, \bar\xi^{\dot\alpha}\}$, $\Theta_i$ is
a constant and we have suppressed the spinor indices. Then $B_i$
can be obtained by applying the following projections,
successively,
\bea\hat\rho_0\;:\;\Phi_i&\longrightarrow&\Phi_{i0}=K_ie^{-iy
F_i\bar{y}-iy\xi_i+i\theta}\,,\nn\\
\hat\rho\;:\;\Phi_{i0}&\longrightarrow&\Phi'_{i0}=K_ie^{-iyF_i
\bar{y}} \Big(e^{-iy\xi_i+i\theta}
+e^{iy\xi_i+i\theta} \Big)\,,\nn\\
\hat\pi\;:\;\Phi'_{i0}&\longrightarrow&B_i=K_ie^{-iyF_i\bar{y}}
\Big(e^{-iy\xi_i+i\theta} +e^{iy\xi_i+i\theta}+e^{-i\bar{y}
\bar\xi_i-i\theta} +e^{i\bar{y}\bar\xi_i-i\theta}\Big)\,.
\label{proj.Bi}\eea
For $B'_i$, we have
\bea B'_i&=&K_i(1-iyF_i \bar{y})\Phi_i\Big|_{\Xi_i=\Theta_i=0}
=K_i\Big(1-\frac{i}2Yf_iY\Big)\Phi_i\Big|_{\Xi_i=\Theta_i=0}\nn\\
&=&K_i\Big(1-\frac{i}2\pd_\Xi f\pd_\Xi\Big)_i\Phi_i\Big|_{\Xi_i
=\Theta_i=0}\,,\label{proj.BiPrime}\\
\pd_\Xi f\pd_\Xi&=&\pd_{\Xi_A} f_A^{~B}\pd_{\Xi^B}\,,\quad
\pd_{\Xi_A}(Y\Xi)=Y^A\,,\quad\pd_{\Xi^A}(Y\Xi)=-Y_A\,.\eea

{\bf 3.} Since both $B_i$ and $B'_i$ can be obtained from $\Phi_i$
either by projection or by an operator that is irrelevant for the
star-product, we can firstly calculate ($\td\Phi =\delta\star
\Phi\star\delta$)
\bea Z_n&=&Tr(\Phi_1\star\delta\star\cdots
\star\Phi_n\star\delta)\nn\\
&=&\left\{\begin{matrix}\Phi_1\star\td\Phi_2\star\cdots\star
\Phi_{n-1}\star\td\Phi_n\Big|_{Y=0}&:&~n{\rm~even}\,,\cr
\int\frac{dy}{2\pi}\Phi_1\star\td\Phi_2\star\cdots
\star\Phi_{n-2}\star\td\Phi_{n-1}\star\Phi_n\Big|_{\bar{y}
=0}&:&n{\rm~odd}\,,\end{matrix}\right.\label{def.trace}\eea
and then apply the above operations to recover $\cZ_n$.

In the following subsections, we work backwards along the steps
outlined here.

\subsection{The building block of generating functions}

In this subsection, we firstly calculate $Z_n$ in
(\ref{def.trace}). Let us note that our computations are $sp(2M)$ covariant although we need the specialization to $sp(4)$ only.

Given (\ref{def.Phi}), we find that
\bea \Phi_1\star\Phi_2&=&\frac{K_1K_2}{\sqrt{|1+f_1f_2|}}
e^{-\frac{i}2Y(f_1\circ f_2)Y-iY(\Xi_1\circ\Xi_2)-\frac{i}2
(\Theta_1\circ\Theta_2)}\,,\nn\\
f_1\circ f_2&=&(2+f_2-f_1)(f_1+f_2)^{-1}\,,\nn\\
\Xi_1\circ \Xi_2&=&\frac12(1+f_1\circ f_2)\Xi_1+\frac12
(1-f_1\circ f_2)\Xi_2\,,\nn\\
\Theta_1\circ\Theta_2&=&-\frac18\Xi_1(f_1\circ f_2+f_2\circ
f_1)\Xi_1-\frac14\Xi_1(1+f_2\circ f_1)\Xi_2\nn\\
&&-\frac18\Xi_2(f_1\circ f_2+f_2\circ f_1)\Xi_2+\frac14
\Xi_2(1+f_1\circ f_2)\Xi_1\nn\\
&&+\Theta_1+\Theta_2\,.\label{def.PhiStarPhi}\eea
In the special case $\breve{f}=f$ and $f^2=1$, one can find the
following useful properties
\bea(f_1+\cdots+f_n)^2=\sqrt{|f_1+\cdots+f_n|}\;\,,\quad
(f_1\circ f_2)\breve~=f_1\circ f_2\,,\nn\\
f_1\circ(f_2\circ f_3)=(f_1\circ f_2)\circ f_3 =f_1\circ
f_2\circ f_3=f_1\circ f_3\,,\nn\\
\quad\Longrightarrow\quad f_1\circ\cdots\circ f_n=f_1\circ f_n\,,\nn\\
(f_1\circ f_2)(f_1\circ f_3)=1+f_1\circ f_2-f_1\circ f_3\,,\nn\\
(f_1\circ f_3)(f_2\circ f_3)=1-f_1\circ f_3+f_2\circ f_3\,,\nn\\
\Xi_1\circ(\Xi_2\circ\Xi_3)=(\Xi_1\circ\Xi_2)\circ\Xi_3 =\Xi_1
\circ\Xi_2\circ\Xi_3=\Xi_1\circ\Xi_3\,,\nn\\
\Longrightarrow\quad\Xi_1\circ\cdots\circ\Xi_n=\Xi_1\circ\Xi_n\,,\nn\\
\Theta_1\circ(\Theta_2\circ\Theta_3)=(\Theta_1\circ\Theta_2)
\circ\Theta_3=\Theta_1\circ\Theta_2\circ\Theta_3\,,\eea
where most relations were already known in \cite{Didenko:2012tv}.
Note, in particular, that the $\circ$-product for $f_i$ and
$\Xi_i$ are ``forgetful". Using such properties, it is easy to
write down the general result
\bea\Phi_1\star\Phi_2\cdots\star\Phi_n&=&N_n\exp\Big\{-\frac{i}2
Y(f_1\circ f_n)Y-iY(\Xi_1\circ\Xi_n)+i(\Theta_1\circ
\cdots\circ\Theta_n)\Big\}\,,\nn\\
N_n&=&2^{2-n}\prod_{i=1}^n\frac{K_{i+1}}{|1+f_if_{i+1}|^{1/4}}
\,,\nn\\
\Theta_1\circ\cdots\circ\Theta_n&=&-\frac18\sum_{i=1}^{n}\Big[
\Xi_i(f_{i+1}\circ f_i+f_i\circ f_{i-1})\Xi_i+2\Xi_i(1+f_{i+1}
\circ f_i)\Xi_{i+1}\Big] +\sum_{i=1}^n\Theta_i\,,
\label{rst.PPPs}\eea
where for any given $n$, we have defined $K_{n+1}=K_1$, $f_{n+1}
=f_1$, $f_0=f_n$ and $\Xi_{n+1}=-\Xi_1$.

Note (\ref{def.trace}) also contains $\td\Phi_i$'s. This can be
easily taken into account by noticing that, to go from $\Phi_i$ to
$\td\Phi_i$, one only need to do the following replacement in
(\ref{rst.PPPs}),
\bea f_i&\longrightarrow&\td{f}_i=I'f_iI'=-f_i\,,\nn\\
\Xi_i&\longrightarrow&\td\Xi_i=I'\Xi_i\,,\quad
I'=\left(\begin{matrix}-1&\cr&1\end{matrix}\right)\otimes
\one_2\,.\eea
The constant $\Theta_i$ is the same for $\Phi_i$ and $\td\Phi_i$.
Plug these results into (\ref{def.trace}) and (\ref{rst.PPPs}),
one can find that
\bea Z_n&=&2^{2-2n}\prod_{i=1}^n\frac{e^{i[\hat{Q}^i
+\hat{P}_{i(i+1)}+\Theta_i]}}{x_{i(i+1)}}\,,
\quad\hat{Q}^i\equiv\hat{Q}^i_{(i+1)(i-1)}\,,\nn\\
\hat{Q}^i_{jk}&=&\Xi_i\cQ^i_{jk}\Xi_i\,,\quad
\cQ^i_{jk}=-\frac18(\td{f}_{j}\circ f_i
+f_i\circ\td{f}_{k})\,,\nn\\
\hat{P}_{ij}&=&\Xi_i\cP_{ij}\Xi_{j}\,,\quad
\cP_{ij}=-\frac14(1+\td{f}_{j}\circ f_i)I'\,. \label{def.Zn}\eea
For later convenience, let's introduce $\Xi_0=\Xi_n$. Let's also
absorb the minus sign in $\Xi_{n+1}$ into $\cP_{n(n+1)}=\cP_{01}
=\cP_{n1}$, so that
\be\Xi_{n+1}=\Xi_1\,,\quad \cP_{n1}=\frac14(1+\td{f}_1\circ
f_n)I'\,.\ee

Now suppose that the $j$'th through $k$'th nods are all $\Delta=2$
scalars, then from (\ref{proj.BiPrime})
\bea Z_n^{j,k}&=&\prod_{i=j}^{k}K_i\Big(1-\frac{i}2\pd_\Xi
f\pd_\Xi\Big)_iZ_n\Big|_{\Xi_j=\cdots=\Xi_{k}=0}\,,\nn\\
&=&\hat{R}_{(j-1)(k+1)}Z_n\Big|_{\Xi_j=\cdots=\Xi_{k}=0}\,,\nn\\
\hat{R}_{(j-1)(k+1)}&=&\Xi_{j-1}\cR_{(j-1)(k+1)}\Xi_{k+1}\,,\nn\\
\cR_{(j-1)(k+1)}&=&i^{k-j+2}\cP_{(j-1)j} \hat{f}_j\cP_{j(j+1)}
\cdots\hat{f}_{k}\cP_{k(k+1)}\,,\label{def.ZnPrime}\eea
where $\hat{f}_i$ is defined in (\ref{prop.fi}). In deriving this
result, we have assumed that not all points are $\Delta=2$
scalars. For the case when all $n$ points are $\Delta=2$ scalars,
we note
\bea Z_n^{2,n}&=&\Xi_1\cR_{1(n+1)}\Xi_1e^{\hat{Q}^1}2^{2-2n}
\prod_{i=1}^n \frac{e^{i\Theta_i}}{x_{i(i+1)}}\,,\nn\\
\Longrightarrow\quad Z_n^{1,n} &=&K_1\Big(1-\frac{i}2\pd_\Xi
f\pd_\Xi\Big)_1Z_n^{2,n}\Big|_{\Xi_1=0}\nn\\
&=&tr[\hat{f}_1\cR_{1(n+1)}]2^{2-2n}\prod_{i=1}^n
\frac{e^{i\Theta_i}}{x_{i(i+1)}}\,.\eea

\subsection{Projections and correlation functions}\label{subsec.proj}

Given $Z_n$ and $Z_n^{j,k}$ we can now impose the projections
defined in (\ref{proj.Bi}).

Firstly note $\xi=-F\bar\xi$, with which one can obtain
\bea Q^i&\equiv&\hat{Q}^i(\xi_i)=\hat{Q}^i(\bar\xi_i)\,,\nn\\
P_{ij}&\equiv&\hat{P}_{ij}(\xi_i,\xi_j)=\hat{P}_{ij}(\bar\xi_i,
\xi_j)=-\hat{P}_{ij}(\xi_i,\bar\xi_j)=-\hat{P}_{ij}(\bar\xi_i,
\bar\xi_j)\,,\nn\\
R_{jk}&\equiv&\hat{R}_{jk}(\xi_i,\xi_j)=\hat{R}_{jk}(\bar\xi_i,
\xi_j)=-\hat{R}_{jk}(\xi_i,\bar\xi_j)=-\hat{R}_{jk}(\bar\xi_i,
\bar\xi_j)\,,\label{defpropPQR}\eea
where $\hat{Q}^i(\xi_i)=\hat{Q}^i|_{\bar\xi_i=0}$, $\hat{P}_{ij}
(\xi_i,\xi_j)=\hat{P}_{ij}|_{\bar\xi_i=\bar\xi_j=0}$ and so on.
Now using (\ref{proj.Bi}), we find
\bea\hat\rho_0Z_n&=&2^{2-2n}\prod_{i=1}^n\frac{e^{i[ Q^i+
P_{i(i+1)}+\theta]}}{x_{i(i+1)}}\,,\nn\\
\hat\rho\hat\rho_0 Z_n&=&2^{2-n}\prod_{i=1}^n \frac{e^{i( Q^i
+\theta)}}{x_{i(i+1)}}\Big[\prod_i^n\cos
 P_{i(i+1)}+i^n\prod_i^n\sin P_{i(i+1)}\Big]\,,\nn\\
\cZ_n=\hat\pi\hat\rho\hat\rho_0 Z_n&=&4\prod_{i=1}^n\frac{
e^{iQ^i}}{x_{i(i+1)}}\Big[\cos^n\theta\prod_i^n\cos P_{i
(i+1)}+\sin^n\theta\prod_i^n\sin P_{i(i+1)}\Big]\,,\nn\\
\hat\rho_0Z^{j,k}_n&=&2^{2-2n}R_{(j-1)(k+1)}\prod_{i=1}^n
\frac{e^{i[Q^i+ P_{i(i+1)}+i\theta]}}{x_{i(i+1)}}
\Big|_{\Xi_j=\cdots=\Xi_{k}=0}\,,\nn\\
\hat\rho\hat\rho_0 Z_n^{j,k}&=&2^{1-n-k+j}i^{n-(k-j+2)}
 R_{(j-1) (k+1)}\prod_{i=1}^n\frac{e^{i( Q^i +\theta)}}{
x_{i(i+1)}}{\prod}'_{i'}\sin P_{i'(i'+1)}\,,\nn\\
\cZ_n^{j,k}=\hat\pi\hat\rho\hat\rho_0Z_n^{j,k}&=&-\frac{i}{
4^{k-j}} (-\sin\theta)^{n-(k-j+1)}R_{(j-1)(k+1)}\prod_{i=1}^n
\frac{e^{iQ^i}}{x_{i(i+1)}}{\prod}'_{i'}\sin P_{i'(i'+1)}\,,\eea
where ${\prod}'_{i'}$ goes over all points for which $P_{i'(i'+1)}
\neq0$ and one has to set $\Xi_j=\cdots=\Xi_{k}=0$ for
$Z_n^{j,k}$.

Our last step is to obtain $V_n$. This can be achieved by applying the permutation
group, $\mathbb{S}_n$, on $\cZ_n$ or $\cZ_n^{j,k}$. Inside
$\mathbb{S}_n$ one can firstly consider the dihedral subgroup,
$D_n$, in which each of the reflections $\hat{s}$ acts as
\bea\hat{s}(Q^i)&=&- Q^i\,,\quad \hat{s}(P_{ij})
=-P_{ij}\,,\quad\hat{s}( R_{jk})=-R_{jk}\,,\nn\\
\Longrightarrow\quad D_n\cZ_n&=&8n\Big[\cos^n\theta\cos
Q\prod_i^n\frac{\cos P_{i(i+1)}}{x_{i(i+1)}}\nn\\
&&\quad+\sin^n\theta f_n( Q)\prod_i^n\frac{\sin P_{i(i+1)}}{
x_{i(i+1)}}\Big]\,,\nn\\
D_n\cZ_n^{j,k}&=&-i\frac{
2n}{4^{k-j}}(-\sin\theta)^{n-(k-j+1)} R_{(j-1)(k+1)}\nn\\
&&\times\prod_{i=1}^n\frac1{x_{i(i+1)}}{\prod}'_{i'}
\sin P_{i'(i'+1)}\nn\\
&&\times f_{n-(k-j+1)}( Q)\,,\eea
where $ Q=\sum_{i=1}^n Q^i$, $f_n(x)=\cos x$ for $n$ even and
$f_n(x)=i\sin x$ for $n$ odd. As a result,
\bea \mathbb{S}_n\cZ_n&=&4\sum_{\mathbb{S}_n}\Big[\cos^n
\theta\cos Q\prod_i^n\frac{\cos P_{i(i+1)}}{x_{i(i+1)}}\nn\\
&&\qquad+\sin^n\theta f_n( Q)\prod_i^n\frac{\sin P_{i(i+1)}}{
x_{i(i+1)}}\Big]\,,\label{rst.final.Vn}\\
\mathbb{S}_n\cZ_n^{j,k}&=&-\frac{i}{4^{n_0-1}}(-\sin
\theta)^{n-n_0}\sum_{\mathbb{S}_n} R_{(j-1)(k+1)}\nn\\
&&\times f_{n-n_0}(Q)\prod_{i=1}^n\frac1{x_{i(i+1)}}{
\prod}'_{i'}\sin P_{i'(i'+1)}\,,\eea
where $n_0(=k-j+1,$ when $k\geq j)$ is the total number of
$\td{j}_0$ insertions. In the case when all $n$ points are
$\Delta=2$ scalars, we have
\bea \brk{\td{j}_0\cdots\td{j}_0}=\sum_{\mathbb{S}_n}
\cZ_n^{1,n}=\sum_{\mathbb{S}_n}tr[\hat{f}_1\cR_{1(n+1)}]
2^{2-2n}\prod_{i=1}^n \frac1{x_{i(i+1)}}\,,\eea

Note $\mathbb{S}_n\cZ_n=V_n=\brk{j_s\cdots j_s}$ in cases without
a $\Delta=2$ scalar, but $\mathbb{S}_n\cZ_n^{j,k}$ is not
equivalent to $V_n$ in cases with $\Delta=2$ scalar operators. The
reason is that $\cZ_n^{j,k}$ only contains contributions from
cases where all $\Delta=2$ scalars are labelled continuously,
while other possibilities are not included. In the case with several $\Delta=2$
scalars, this means there are mixed sequences such as
\be\cdots j_s\td{j}_0\cdots\td{j}_0j_s\cdots j_s\td{j}_0\cdots
\td{j}_0j_s\cdots j_s\td{j}_0\cdots\td{j}_0j_s\cdots\,.\ee
In this sequence, the $Q$, $P$ and $R$ structures can be read off
as follows,
\bea j_s(i)&\longrightarrow&Q^i\,,\nn\\
j_s(i)j_s(i+1)&\longrightarrow&P_{i(i+1)}\,,\nn\\
j_s(i)\Big({\rm a~sequence~of~}\td{j}_0\Big)j_s(k)
&\longrightarrow&R_{ik}\,,\label{def.QPRseq}\eea
where $j_s(i)$ means that $j_s$ is on the $i$'th position in the
sequence, and similarly $\td{j}_0(i)$ means that $\td{j}_0$ is on
the $i$'th position in the sequence. Using $\{\cdots\}$ to denote
the above sequence, and using $\cZ_n^{\{\cdots\}}$ to denote the
corresponding generating function, we find
\bea\mathbb{S}_n\cZ_n^{\{\cdots\}}&=&\ds\frac{(i\sin\theta)^{n
-n_0}}{4^{n_0-1}}\sum_{\mathbb{S}_n}f_{n-n_0}(Q)\prod_{i=1}^n
\frac1{x_{i(i+1)}}\Upsilon_{\{\cdots\}}^{-1}\prod R\prod
(i\sin P)\,,\nn\\
V_n({\rm with~}\td{j}_0)&=&\sum_{\{\cdots\}}\mathbb{S}_n
\cZ_n^{\{\cdots\}}\,, \label{rst.final.Vnj0}\eea
where the last two $\prod$ in $\mathbb{S}_n\cZ_n^{\{\cdots\}}$ go
over all non-vanishing $R$'s and $P$'s, which are determined
according to (\ref{def.QPRseq}); and $\Upsilon_{\{\cdots\}}$
is the operator that takes the canonical sequence
$\{\td{j}_0\cdots\td{j}_0j_s\cdots j_s\}$ to the particular
sequence $\{\cdots\}$ that is involved. As an example, let's note
\bea \brk{\td{j}_0\td{j}_0j_sj_s}&=&\mathbb{S}_n\cZ_4^{\ds
\{\td{j}_0\td{j}_0j_sj_s\}}+\mathbb{S}_n\cZ_4^{\ds\{\td{j}_0
j_s\td{j}_0j_s\}}\nn\\
&=&-\frac14\sin^2\theta\sum_{\mathbb{S}_n}{}\Big[\cos(Q^3+Q^4)\frac{iR_{03}\sin P_{34}}{
x_{12}x_{23}x_{34}x_{41}}+\Upsilon_{\{\td{j}_0
j_s\td{j}_0j_s\}}^{-1} \cos(Q^2+Q^4)\frac{R_{02}R_{24}}{x_{12}x_{23}x_{34}x_{41}}\Big]\,,\nn\eea
where $\Upsilon_{\{\td{j}_0j_s\td{j}_0j_s\}}$ is the operator that
takes the sequence $\{\td{j}_0\td{j}_0j_sj_s\}$ to
$\{\td{j}_0j_s\td{j}_0j_s\}$ (it acts on all of $x_i$ and $\eta_i$), c.f. the third correlator of \eqref{exmp.4pt}.

As a side remark, we have explicit $i$-factors floating around in
our final results (\ref{rst.final.Vn}) and (\ref{rst.final.Vnj0}),
and also inside $ R_{ij}$ as in (\ref{def.PQR}). The appearance of
these $i$-factors is due to the fact that we have neglected some
extra phase factors that naturally arise in
(\ref{def.PhiStarPhi}), as is shown in \cite{Gelfond:2013xt}.
Consistency requires that the result must be hermitian and all
these $i$-factors must cancel when all phase factors are taken
into account properly. Our main objective here is to obtain the
basic structure of correlation functions since an overall factors
are undetermined within our approach.

\section{Conclusion}\label{sec:conclusions}

In this note we complete the calculation of $n$-point correlation
functions of conserved currents in unbroken $4d$ Vasiliev theory
initially carried out for $\Delta=s+1$ operators in
\cite{Didenko:2012tv}. The missing link that we deal with in our
paper is the correlation functions that contain $\Delta=2$ scalar
operator $\tilde{j}_0$ \eqref{corr}. The obtained results are in
agreement with Maldacena-Zhiboedov theorem, with three and some
four-point calculations performed using different methods
\cite{Giombi:2009wh, Giombi:2010vg, Giombi:2011rz, Maldacena:2011jn, Colombo:2012jx} and in agreement with very
recent calculation \cite{Gelfond:2013xt} where $n$-point functions
were reproduced from the current operator algebra. Our method is
applicable only when HS symmetry is unbroken and it is promising to look whether it can
be improved to the case when HS symmetry is broken. The trace formula \eqref{allcorrfunc}
determines the correlation functions up to overall coefficients and
makes the calculations very simple. The price to pay for this
simplicity is that we only reproduce connected part of correlation
functions unlike the complete result of \cite{Gelfond:2013xt}. The
great advantage of the proposed method is its manifest conformal
and HS invariance. HS boundary-to-bulk propagators as well as
conformal structures arise in a coordinate independent way. All
have beautiful interpretation in HS algebra, the former correspond
to projectors in star-product algebra, while the latter appear
naturally through the induced $\circ$-product defined on a
projector space. The use of coordinates is thus the matter of
presenting the results to make contact with the available ones in
the literature.

Our method can be straightforwardly generalized to any dimension
although in higher dimensions HS algebra has no longer simple
realization similar to lower $d$ spinorial Weyl algebras. That the
analogous $d$-dimensional calculation is not going
to be simple at all has been already demonstrated at the level of
HS propagators in \cite{Didenko:2012vh} unless the induced product on the space
of projectors is understood\footnote{Three-point functions have been recently found in \cite{Zhiboedov:2012bm}.}. On the other hand, the
spinorial route to HS algebras naturally extends to simplectic HS
algebras.

\section*{Acknowledgement}
We would like to thank Alexander Zhiboedov and Mikhail Vasiliev
for valuable discussions and comments. The work of J.M and E.S was
supported by the Alexander von Humboldt-Foundation. The work of
V.D was supported in part by the grant of the Dynasty Foundation.
The work of E.S and V.D was supported in part by RFBR grant
No.11-02-00814, 12-02-31837.

\begin{appendix}
\renewcommand{\theequation}{\Alph{section}.\arabic{equation}}

\section{Technicalities}
\setcounter{equation}{0}\setcounter{section}{1}
\label{secDetails}
The $\Delta=1$ propagator can be constructed from the following,
\bea \Phi_i&=&K_ie^{-iyF_i\bar{y}-iy\xi_i-i\bar{y}\bar\xi_i}\,,
\quad\xi_i=\Pi_i\eta_i\,,\quad\bar\xi_i=\bar\Pi_i\bar\eta_i\,,\nn\\
K_i&=&K(\rho,x;x_i)=\frac\rho{\hat{x}_i^2+\rho^2}\,,\nn\\
F_i&=&F(\rho,x;x_i)=-\Big(2K_i\hat{\bx}_i+i\frac{
\hat{x}_i^2-\rho^2}{\hat{x}_i^2+\rho^2}\Big)\,,\nn\\
\Pi_i&=&\Pi(\rho,x;x_i)=e^{-i\pi/4}K_i\Big(\frac1{\sqrt\rho}
\hat{\bx}_i-i\sqrt\rho\Big)\,,\nn\\
\bar\Pi_i&=&\bar\Pi(\rho,x;x_i)=e^{-i\pi/4}K_i\Big(\frac1{
\sqrt\rho}\hat{\bx}_i+i\sqrt\rho\Big)\,,\label{def.B1.F}\eea
where $\eta$ and $\bar\eta$ are {\it real} constant boundary
polarization vectors, and
\bea \hat{\bx}_i\equiv\bx-\bx_i\,,\quad\bx=\left(\begin{matrix}
-x_2&-x_0-x_1\cr x_0-x_1&x_2\end{matrix}\right)\,,\nn\\
\Longrightarrow\quad x^2=-\det(\bx)=x_0^2-x_1^2-x_2^2\,.\eea
Note the properties
\bea F_i\breve{F}_i=1\,,\quad \breve\Pi_i\Pi_i=\Pi_i\breve\Pi_i
=\breve{\bar\Pi}_i\bar\Pi_i=\bar\Pi_i\breve{\bar\Pi}_i=-iK_i\,,\nn\\
\breve{\bar\Pi}_i=\breve{F}_i \Pi_iF_i\,,\quad \Pi_i^{-1} F_i
\breve{\bar\Pi}^{-1}=\Pi_i^{-2}\breve{F}_i^{-1}=\frac1{K_i}\,.
\label{prop.FandPi}\eea
From $\xi=-F\bar\xi$, one can derive that $\bar\xi^\ast=\xi$ and
$\bar\eta=-i\eta$. It will be convenient to write
\bea\Phi_i=K_ie^{-\frac{i}2Yf_iY-iY\Xi_i}\,,\quad Y=\left(
\begin{matrix}y\cr\bar{y}\end{matrix}\right)\,,\quad\Xi_i=
\left(\begin{matrix}\xi_i\cr\bar\xi_i\end{matrix}\right)
=\Gamma_i\Xi_i^0\,,\nn\\
f_i=\left(\begin{matrix}&F_i\cr\breve{F}_i&\end{matrix}\right)
\,,\quad \Gamma_i=\left(\begin{matrix}\Pi_i&\cr&\bar\Pi_i
\end{matrix}\right)\,,\quad\Xi_i^0=\left(\begin{matrix}\eta_i
\cr\bar\eta_i\end{matrix}\right)\,.\label{def.B1}\eea
From (\ref{prop.FandPi}), one has
\be\breve{f}_i=f_i\,,\quad f_i^2=1\,,\quad\Gamma_i^{-1}
\hat{f}_i\breve\Gamma_i^{-1}=i\left(\begin{matrix}&-1\cr1
&\end{matrix}\right)\otimes\one_2\,,\quad \hat{f}_i\equiv
-iK_if_i\,,\label{prop.fi}\ee
where $\one_2$ means 2-dimensional unit matrix.

From (\ref{def.Zn}) and (\ref{def.ZnPrime}), we find that
$\breve{\cQ}^i_{jk}=\cQ^i_{jk}$, $tr(f_i\cQ^i_{jk})=-1$ and
\bea\breve\Gamma_i\cQ^i_{jk}\Gamma_i&=&\frac18\left(\begin{matrix}
\bx_{ij}^{-1}-\bx_{ik}^{-1}&\cdots\cr \cdots&-(\bx_{ij}^{-1}
-\bx_{ik}^{-1})\end{matrix}\right)\,,\nn\\
\breve\Gamma_i\cP_{ij}\Gamma_j&=&-\frac14\left(\begin{matrix}1&-i
\cr i&1\end{matrix}\right)\otimes \bx_{ij}^{-1}\,,\nn\\
\breve\Gamma_{j-1}\cR_{(j-1)(k+1)}\Gamma_{k+1}&=&\frac{c}{
2(2i)^{k-j+2}}\left(\begin{matrix}1&-i\cr i&1\end{matrix}\right)
\otimes\Big(\prod_{i=j-1}^{k}\bx_{i(i+1)}^{-1}\Big)\,,
\label{prop.QPR}\eea
where $c=1$ if $\bx_{n1}^{-1}$ is not involved while $c=-1$ if
otherwise. As a result,
\bea Q^i&=&-\frac18\eta_i[\bx_{i(i+1)}^{-1}
-\bx_{i(i-1)}^{-1}]\eta_i\,,\nn\\
P_{ij}&=&\frac{c}4\eta_i\bx_{ij}^{-1}\eta_j\,,\nn\\
R_{(j-1)(k+1)}&=&-\frac{c}{2(2i)^{k-j+2}}\eta_{j-1}
\Big(\prod_{i=j-1}^{k}\bx_{i(i+1)}^{-1}\Big)
\eta_{k+1}\,.\label{def.PQR}\eea
Note that the off diagonal elements of $\breve\Gamma_i\cQ^i_{jk}
\Gamma_i$ do not contribute to these structures. One can further
find that
\bea Z_n^{1,n}&=&tr\Big[-\frac1{(2i)^n}\left(\begin{matrix} 1&
-i\cr i&1\end{matrix}\right)\otimes\Big(\bx_{12}^{-1}\bx_{23}^{
-1} \cdots \bx_{n1}^{-1}\Big)\Big]2^{2-2n}\prod_{i=1}^n
\frac1{x_{i(i+1)}}\nn\\
&=&-8\ds\frac{tr(\bx_{12}^{-1}\bx_{23}^{-1}\cdots
\bx_{n1}^{-1})}{(8i)^nx_{12}x_{23}\cdots x_{n1}}\,.\eea

Under the inversion, $\bx_i\to\bx_i^{-1}$ and $\Xi_i^0\to
\bx_i^{-1}\Xi_i^0$. One has
\bea\bx_{ij}&\longrightarrow&\bx_i^{-1}-\bx_j^{-1}=-\bx_i^{-1}
\bx_{ij}\bx_j^{-1}=-\frac{\bx_i\bx_{ij}\bx_j}{x_i^2x_j^2}\,,\nn\\
x_{ij}^2=-|\bx_{ij}|&\longrightarrow&-|\bx_i^{-1}\bx_{ij}\bx_j^{-1}|
=-\frac{|\bx_{ij}|}{|\bx_i||\bx_j|}=\frac{x_{ij}^2}{x_i^2x_j^2}\,,\nn\\
\bx_{ij}^{-1}=\frac{\bx_{ij}}{x_{ij}^2}&\longrightarrow&-\frac{
\bx_i\bx_{ij}\bx_j}{x_i^2x_j^2}\Big/\frac{x_{ij}^2}{x_i^2x_j^2}
=-\bx_i\bx_{ij}^{-1}\bx_j\,,\nn\\
\#_k&\longrightarrow&(-\Xi_{i_1}^0\bx_{i_1}^{-1})(-\bx_{i_1}
\bx_{i_1i_2}^{-1}\bx_{i_2i_3}\bx_{i_3i_4}^{-1}\cdots\bx_{i_{k-2}
i_{k-1}}\bx_{i_{k-1}i_k}^{-1} \Xi_{i_k}^0)=\#_k\,,
\label{trans.inversion}\eea
where $\#_k=\Xi_{i_1}^0\bx_{i_1i_2}^{-1}\bx_{i_2i_3} \bx_{i_3
i_4}^{-1} \cdots\bx_{i_{k-2}i_{k-1}}\bx_{i_{k-1}i_k}^{-1} \Xi_{
i_k}^0$. This implies, in particular, that $ P_{ij}$ and $ Q^i$
are conformal invariants. One can then check that under
(\ref{trans.inversion}),
\bea Z_n&\longrightarrow&\Big(\prod_i^nx_i^2\Big)Z_n\,,\nn\\
Z_n^{j,k}&\longrightarrow&\Big(\prod_i^nx_i^2
\prod_{j'=j}^{k}x_{j'}^2\Big)Z_n^{j,k}\,,\nn\\
Z_n^{1,n}&\longrightarrow&\Big(\prod_i^nx_i^4\Big)Z_n^{1,n}\,,\eea
which all transform as expected.

\end{appendix}

\providecommand{\href}[2]{#2}\begingroup\raggedright\endgroup

\end{document}